\documentclass[aps,prd,reprint,groupedaddress,floatfix]{revtex4-1}

\usepackage[colorlinks=true,citecolor=blue,urlcolor=magenta,breaklinks]{hyperref}
\usepackage{natbib}
\usepackage{graphicx}
\usepackage[caption=false]{subfig}
\usepackage{hyperref}
\usepackage{wasysym}
\usepackage{color}
\usepackage{amsfonts}

\definecolor{rgbgreen}{rgb}{0, 0.5, 0}

\definecolor{rgbmagenta}{rgb}{0.5, 0, 0.5}

\begin{document}

\title{Asteroseismic constraints on Asymmetric Dark Matter: \\ Light particles with an effective spin-dependent coupling}

\author{Andr\'{e} Martins}
\email{andre.mp.martins@gmail.com}
\author{Il\'{i}dio Lopes}
\email{ilidio.lopes@ist.utl.pt}
\affiliation{CENTRA, Departamento de F\'{i}sica, Instituto Superior T\'{e}cnico, Universidade de Lisboa, Av. Rovisco Pais 1, 1049 Lisboa, Portugal}
\author{Jordi Casanellas}
\email{jordicasa@gmail.com}
\affiliation{Max Planck Institut f\"{u}r Gravitationsphysik (Albert-Einstein-Institut), D-14476 Potsdam, Germany}

\date{\today}

\begin{abstract}
So far, direct detection searches have come empty handed in their quest for Dark Matter (DM). Meanwhile, asteroseismology arises as a complementary tool to study DM, as its accumulation in a star can enhance energy transport, by providing a conduction mechanism, producing significant changes in the stellar structure during the course of the star's evolution. The stellar core, particularly affected by the presence of DM, can be investigated through precise asteroseismic diagnostics. We modelled three stars including DM energy transport: the Sun, a slightly less massive and much older star, KIC 7871531 ($0.85 \, \text{M}_{\astrosun}$, $9.41 \, \text{Gyr}$), and a more massive and younger one, KIC 8379927 ($1.12 \, \text{M}_{\astrosun}$, $1.82 \, \text{Gyr}$). We considered both the case of Weakly Interactive Massive Particles, albeit with a low annihilation, and the case of Asymmetric DM for which the number of trapped particles in the star can be much greater. By analysing these models with asteroseismic separation ratios weighted towards the core, we found indications limiting the effective spin-dependent DM-proton coupling for masses of a few GeV. This independent result is very close to the most recent and most stringent direct detection DM constraints.
\end{abstract}

\maketitle

\section{Introduction}

Elusive DM has so far evaded direct, indirect, collider and astrophysical searches. For now, the experiments PICO-2L and PICASSO have constrained the effective DM-proton spin-dependent (SD) cross section $\sigma_{\chi n}^{SD}$ to values just below $10^{-37} \, \text{cm}^{2}$ for a DM particle with a mass $m_{\chi} = 5 \, \text{GeV}$ \cite{Olive2014c}. If DM particles are thermal relics such as Weakly Interactive Massive Particles (WIMPs), then the natural scale for the (thermally averaged) annihilation cross section $\left\langle \sigma_{A} v \right\rangle = 3 \times 10^{-26} \, \text{cm}^{3}/\text{s}$ \cite{Steigman2012}. However, the latest Fermi-LAT dwarf satellite galaxy observations have constrained the annihilation cross section to values just slightly lower than that natural scale \cite{Ahnen2016}. On the other hand, in the Asymmetric DM (ADM) scenario, present-day DM annihilation is negligible. Additionally, in general DM particles may also interact with each other without annihilating. From an analysis of colliding galaxy clusters, \citet{Harvey2015} recently set robust limits which constrain the self-interaction cross section to $\sigma_{\chi \chi} / m_{\chi} \lesssim 8.3 \times 10^{-25} \, \text{cm}^{2}/\text{GeV}$.

Meanwhile, helioseismology has been used as a complementary tool to constrain the properties of DM \cite{Turck-Chieze2012}. This is possible because DM particles accumulating in a star transport energy by conduction, affecting its stellar structure, in particular in the stellar core.  The presence of WIMPs has been shown to significantly alter the local luminosity and sound speed in the Sun, allowing constraints to be set through a comparison between helioseismic data and solar models including DM \cite{Lopes2002c}. This approach has been extended by \citet{Lopes2002d} to include constraints from solar neutrinos and from solar gravity modes by \citet{Turck-Chieze2012c}. WIMPs with an annihilation cross section close to the natural scale do not accumulate in large enough numbers inside the Sun to produce an impact incompatible with the observational data. On the other hand, \citet{Frandsen2010} showed that accumulation can be greatly enhanced for self-interacting ADM, thus producing a significant impact in the Sun. Considering a WIMP annihilation cross section several orders of magnitude below the natural scale also has this effect. WIMP-like ADM, for which ADM is emulated by WIMPs with a very low annihilation rate, has been investigated, with constraints having been set by \citet{Cumberbatch2010,Taoso2010} and \citet{Lopes2012}. Furthermore, \citet{Lopes2014} studied an ADM scenario with long-range DM-baryon interactions. Recently, \citet{Vincent2015} showed that a solution including $q^{2}$ momentum-dependent ADM reasonably fits the data with $\sigma_{\chi n} = 10^{-37} \, ( q / 40 \, \text{MeV} )^{2} \, \text{cm}^{2}$ and a low $m_{\chi} = 3 \, \text{GeV}$, somewhat below the typical effective-interaction evaporation threshold \citep{Vincent2015a}.

Adding to this, the Sun is no longer the only star we can conduct these studies with. The COROT \cite{Baglin2006,Michel2008} and \textit{Kepler} \cite{Gilliland2010} missions revolutionized asteroseismology by detecting stellar oscillations with a remarkable precision for thousands of solar-like and red giant stars \citep{Chaplin2013a}. We are now in a position to take advantage of that contribution, by using those stars as laboratories for fundamental physics. \citet{Casanellas2011} suggested that the use of diagnostics from stellar oscillations could be used to constrain the properties of DM. In a follow-up, \citet{Casanellas2013} reported the first asteroseismic constraints for WIMP-like ADM from solar-like stars.

Parallel to this situation there is a long-running predicament in astrophysics, the solar composition problem. The issue is the discrepancy between the solar structure inferred from helioseismology and that obtained in Standard Solar Models (SSMs) inputing the most up-to-date photospheric abundances \citep{Serenelli2009}. The solar composition problem is relevant not only for solar models, but also for any stellar model, since the abundances in the Sun which are generally assumed to be similar to other solar-type stars are a crucial input for any stellar model. The solar metallicity worked out almost three decades ago has since been revised to lower values and more recently it was brought slightly up to a reliable present-day $Z/X = 0.0181$ \cite{Asplund2009d,Scott2014,Scott2014a,Grevesse2014}. Yet, the problem persists with possible solutions ranging from more accurate spectroscopic analysis and radiative opacities, to an enhanced neon abundance, to more accurate stellar modelling \citep{Bergemann2014d}. Interestingly though, the solar composition and the DM problems may not be as parallel as initially thought, with DM possibly, yet unlikely, explaining the difference between the solar structure inferred from helioseismology and that obtained for current SSMs.

We modelled three stars including DM in two scenarios, WIMP-like ADM with
\begin{eqnarray}
\left\langle \sigma_{A} v \right\rangle = 10^{-33} \, \text{cm}^{3}/\text{s} \nonumber
\end{eqnarray}
and ADM with
\begin{eqnarray}
\sigma_{\chi \chi} = 10^{-24} \, \text{cm}^{2}/\text{GeV} . \nonumber
\end{eqnarray}
We concern ourselves exclusively with an effective DM-proton SD interaction cross section and the region of the parameter space explored here is:
\begin{eqnarray}
4 \lesssim m_{\chi} / \text{GeV} \lesssim 15 , \quad 10^{-37} \lesssim \sigma_{\chi n}^{SD} / \text{cm}^{2} \lesssim 10^{-34} \nonumber
\end{eqnarray}
Besides the Sun we modelled the less massive KIC 7871531, with a modelled mass of $0.85 \, \text{M}_{\astrosun}$ and spectral type G5V \citep{Molenda-Zakowicz2013} and the more massive F9IV-V KIC 8379927 with a modelled mass of $1.12 \, \text{M}_{\astrosun}$. 

Asteroseismology has been used before to study the effects of WIMP-like ADM in stars less massive than the Sun \citep{Casanellas2013}. However, this is the first time that asteroseismic signatures of a star less massive than the Sun are used to study self-interacting ADM, hereafter known simply as ADM. Moreover, that same less massive star is also a very old one, with a model age of $9.41 \, \text{Gyr}$, which means that DM accumulates to greater numbers, producing a significantly greater impact.

In section \ref{ADM} we discuss the differences between WIMP and ADM accumulation inside a star, then in section \ref{stars} we revisit how those trapped particles expedite the transport of energy, consequently having an impact on stellar structure during the course of stellar evolution. In section \ref{model} we elaborate on how we picked, modelled and calibrated these particularly appropriate stars. We then proceed to present our results in section \ref{results}, upon which we discuss our findings in section \ref{discussion_conclusions}.

\section{Accumulation of self-interacting Asymmetric DM inside a star} \label{ADM}

\begin{figure*}

\subfloat[$\sigma_{\chi \chi} = 10^{-24}, \, 10^{-25} \text{ and } 10^{-26} \, \text{cm}^{2}$ for the solid, dashed and dotted lines, respectively. \label{adm_wimp_xx}]
{\includegraphics[width=0.98\columnwidth]{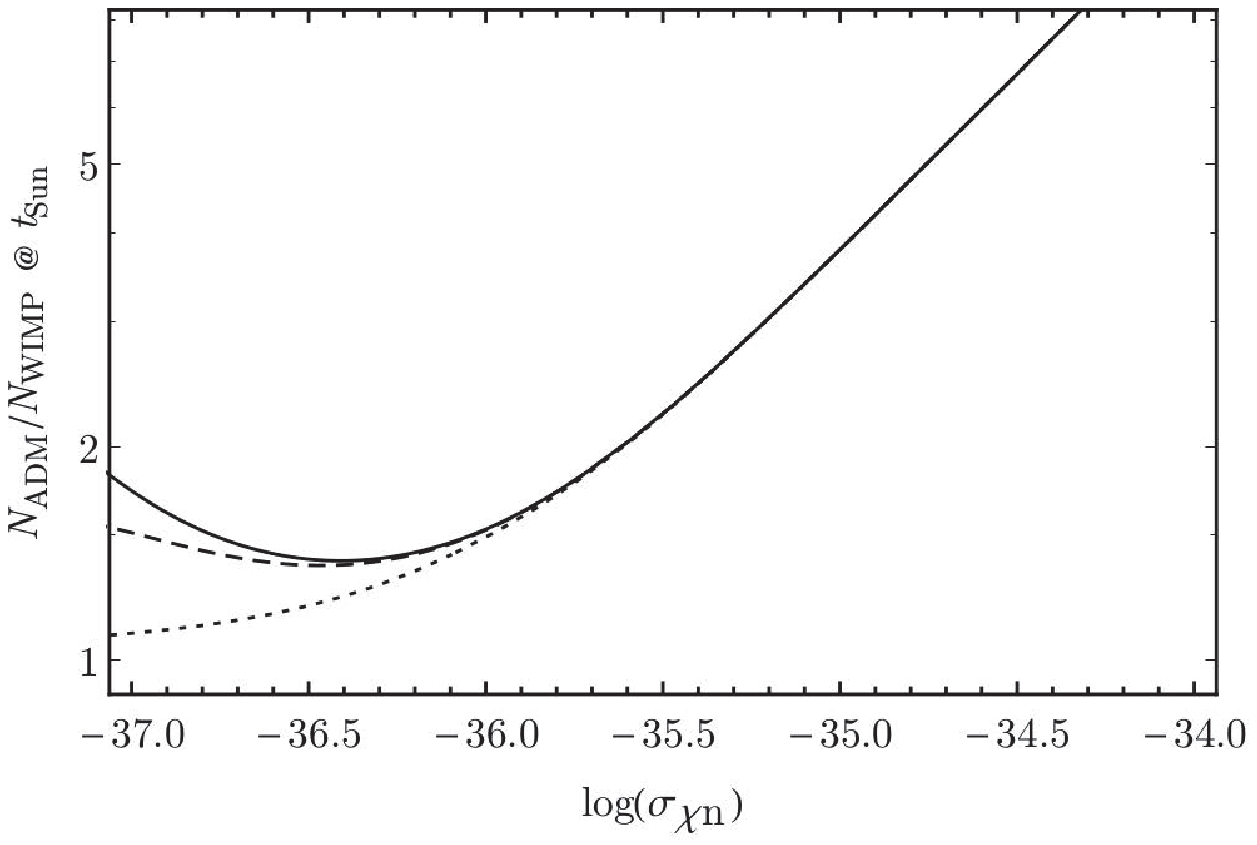}}
~
\subfloat[$\left\langle \sigma_{A} v \right\rangle = 10^{-33}, \, 10^{-30}, \text{ and } 10^{-26}, \, \text{cm}^{3}/\text{s}$ for the solid, dashed and dotted lines, respectively. \label{adm_wimp_an}]
{\includegraphics[width=0.98\columnwidth]{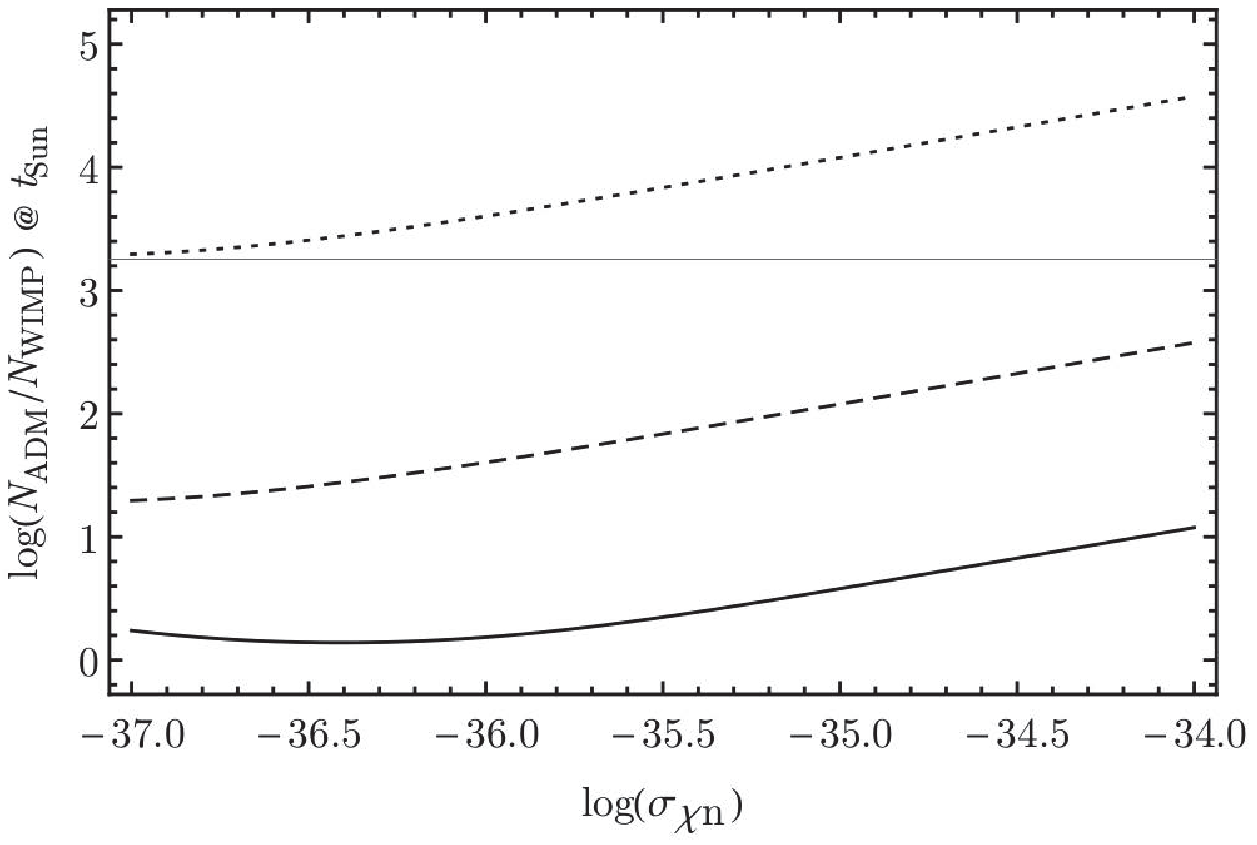}}

\caption{Analytical approximations to the hypothetical present solar value of $N_{ADM}/N_{WIMP}$ as a function of the DM-nucleon interaction cross section. Figure \ref{adm_wimp_xx} is for $\left\langle \sigma_{A} v \right\rangle = 10^{-33} \, cm^{3}/s$ and figure \ref{adm_wimp_an}, is actually showing the $log_{10}$ of $N_{ADM}/N_{WIMP}$, for $\sigma_{\chi \chi} = 10^{-24} \, cm^{2}$. \label{adm_wimp}}
\end{figure*}

The fact that present-day baryonic and dark matter densities are of the same order of magnitude lead to the idea of a connection between these components. This connection can be realized by an asymmetry generated in any or both sectors which is then communicated between them, giving rise not only to the baryonic asymmetry, but also to ADM models \citep{Petraki2013,Zurek2014}. In ADM scenarios, today only DM particles would remain, after DM antiparticles vanished by annihilating with the former. Hence, these scenarios contrast with the self-conjugate WIMP picture in that there is no annihilation at present time. Thus, to correctly estimate the number of trapped particles inside a star in the ADM scenario, it is necessary to consider DM self-interaction, as the larger number of already accumulated particles may increase the self-capture rate significantly.

The number of trapped WIMPs inside a star would evolve as
\begin{equation}
N_{WIMP} (t) = \sqrt{ C_{\chi n} / \Gamma_{sa} } \tanh \left( t \sqrt{ C_{\chi n} \Gamma_{sa} } \right) ,
\end{equation}
where $C_{\chi n}$ is the DM capture rate due to DM halo particles scattering off nucleons and $\Gamma_{sa}$ is the self-annihilation rate. For the WIMP picture it is not necessary to consider self-interaction, because DM capture due to scattering off nucleons is much more relevant than self-capture when annihilation is considered. The balance is essentially between the dominant capture mechanism and annihilation.

On the other hand, the number of trapped ADM particles inside a star would evolve as
\begin{equation}
N_{ADM} (t) =  C_{\chi n} / C_{\chi \chi} \left( \exp( C_{\chi \chi} t ) - 1 \right),
\end{equation}
until a geometric limit is reached when the trapped DM becomes optically thick at which point $N_{ADM}$ is driven only linearly with time. Here $C_{\chi \chi}$ is the capture rate due to DM particles in the halo scattering off of other DM particles already trapped inside the Sun. The capture rates $C_{\chi n} \propto \sigma_{\chi n}$ and $C_{\chi \chi} \propto \sigma_{\chi \chi}$ are approximated as in \cite{Zentner2009}. The annihilation rate $\Gamma_{sa} \propto \left\langle \sigma_{A} v \right\rangle$ was given by \cite{Griest1987}. The geometric limit  is reached early on in the life of a star for the parameter space explored in this work.

The capture rates are also determined by the DM halo parameters near the star, by the galactic orbital velocity $v_{*}$ of the object and by the escape velocity at the surface of the target $v_{esc}$ \cite{Zentner2009}. The two stars studied here besides the Sun are located close to it, thus local DM halo parameters are used for all models, namely a density $\rho_{\chi} = 0.38 \, \text{GeV/cm$^{3}$} $ and a velocity dispersion $\bar{v_{\chi}} = 270 \, \text{km/s}$ \cite{Read2014, Choi2014}. The escape velocity $v_{esc}$ is computed for each star, at each time step. All models are also computed with a local Milky Way orbital velocity $v_{*} = 220 \, \text{km/s}$ which can differ, within the same order of magnitude, for the other two stars. However, that difference introduces only a small error for the capture \cite{Lopes2011}.

For comparison of ADM and WIMP accumulation, figures \ref{adm_wimp_xx} and \ref{adm_wimp_an} show analytical approximations to the present solar fraction of the number of trapped DM particles in the ADM scenario relative to the WIMP picture. The number of ADM particles trapped inside the Sun is greater than that of WIMPs by a factor of a few for $\left\langle \sigma_{A} v \right\rangle \sim 10^{-33} \, \text{cm}^{3}/\text{s}$ and by $10^{4}$ relative to the natural scale of annihilation for a thermal relic. Figure \ref{adm_wimp_xx} clearly evidences that, despite the fact that no-annihilation drastically changes DM accumulation, the number of trapped ADM particles is very stiff with respect to the self-capture cross sections. In fact, lowering the value of the self-capture cross section considered here, $\sigma_{\chi \chi} = 10^{-24} \, \text{cm}^{2}$, to $10^{-26} \, \text{cm}^{2}$ reduces the number of trapped particles by less than a factor of $2$ and only for lower values of the DM-nucleon cross section. Moreover, the annihilation cross section considered for WIMPs, $\left\langle \sigma_{A} v \right\rangle \sim 10^{-33} \, \text{cm}^{3}/\text{s}$, is several orders of magnitude lower than the natural scale and would lead to overclosure. We merely adopted the low value for comparison with self-interacting ADM, by emulating the former as WIMP-like ADM. Greater annihilation cross sections do not allow for enough WIMP accumulation in a star to produce a significant effect. In contrast, in the ADM scenario there is no annihilation, so going to unusually low annihilation cross sections is not required to obtain a significant impact on stars, and moreover, for ADM the self-capture mechanism becomes relevant.

A trapped DM particle can evaporate by scattering off a proton and gaining enough energy to escape the gravitational potential of the star. Gould determined the minimum mass a DM particle must have in order to stay in the Sun and not evaporate \citep{Gould1987a,Gould1990}. For the parameter space explored in this work, evaporation in the Sun is essentially irrelevant for masses $m_{\chi} \gtrsim 3.7 \, \text{GeV}$. The evaporation mass for the other stars can be only slightly different \citep{Zentner2011}, so evaporation can be safely neglected as we work for $m_{\chi} > 4 \, \text{GeV}$.

\section{DM energy transport and Asteroseismic diagnostics} \label{stars}

DM provides an additional energy transport mechanism in a star. DM particles can conduct energy from the stellar interior to the outer layers, significantly affecting stellar structure. Consequently, DM impacts stellar oscillations, from which precision diagnostics can be used to explore the properties of DM.

We use \verb|dmp2k| to compute stellar structure and evolution including DM energy transport. \verb|dmp2k| integrates \verb|CESAM2k| \citep{Morel2007} and a collection of routines, some based on \verb|DarkSUSY| \citep{Gondolo2004}. \verb|CESAM2k| calculates 1D quasi-static stellar structure and evolution including diffusion. Similarly to what is described by \cite{Lopes2002c,Scott2009,Lopes2011}, to emulate the effects of DM energy conduction we included an extra transport luminosity
\begin{equation}
L_{trans}(r,t) = \mathfrak{s}(K,r,t) L_{trans, LTE}(r,t) ,
\end{equation}
where $\mathfrak{s}(K,r,t)$ is a suppression factor depending on the Knudsen number $K$ and on the radius $r$ and age $t$. To emulate the non-local energy transport regime due to an isothermal DM distribution, the suppression factor decreases the energy transported by DM particles distributed in Local Thermodynamic Equilibrium (LTE), \\ \\ \\
\begin{eqnarray}
L_{trans, LTE}(r,t) = 4 \pi r^{2} \kappa(r,t) n_{\chi, LTE}(r, t) l(r,t) \times \nonumber \\
\times \left[ \frac{ k_{B} T_{\star}(r,t) }{ m_{\chi} c^{2} } \right]^{1/2} k_{B} \frac{d T_{\star}(r,t)}{dr} ,
\end{eqnarray}
with $\kappa$ the opacity, $n_{\chi, LTE}$ the number density of DM particles in LTE, $l$ the mean free path of these particles and $T_{\star}$ the stellar temperature.

The quasi-static equilibrium problem is solved to a required precision level of $10^{-4}$ using between $500$ and $1000$ mass shells and evolution is computed in about $50$ time steps. The solar photospheric abundances of Asplund, Grevesse, Sauval and Scott, AGSS09ph \citep{Asplund2009,Serenelli2009} are adopted for the chemical composition. The mixing length theory parameter is chosen fixed at $\alpha = 1.8$ without overshoot. The OPAL 2001 tables \citep{Rogers2002} are used for the equation of state and the OPAL+Alexander tables \citep{Rogers1996,Ferguson2005} for the Rosseland mean of the opacities. The NACRE compilation \citep{Angulo1999,Morel1999} is used for the thermonuclear reaction rates.

The oscillation mode frequencies $\nu_{n,\ell}$ for the radial order $n$ and spherical degree $\ell$ are computed from the stellar models using the Aarhus adiabatic pulsation package \citep{Christensen-Dalsgaard2008}. Stellar interior diagnostics can then be determined from the oscillation mode frequencies. Here, we focus on $r_{02}$, the ratio of small to large separations proposed by \cite{Roxburgh2003a},
\begin{equation}
r_{02} (n) = \frac{ \nu_{n,0} - \nu_{n-1,2} }{ \nu_{n,1} - \nu_{n-1,1} } ,
\end{equation}
which is weighted towards the stellar core, where the effects of DM are most relevant. Since $r_{02}$ is independent of the outer layers, it is not significantly affected by inaccurate atmospheric modelling. Our analysis is then based on the statistical test
\begin{equation}
\chi^{2}_{r_{02}} = \sum_{n} \left( \frac{r_{02}^{obs}(n)-r_{02}^{mod}(n)}{\sigma_{r_{02}^{obs}}(n)} \right)^{2} .
\end{equation}
Because $r_{02}(n)$ ratios with consecutive orders share one eigenfrequency, the different ratios used could in principle be correlated. However, by generating samples of the ratios through the sampling of the observed normally distributed eigenfrequencies, we found that in general the correlation is very small, bellow $0.01$. Besides $r_{02}$, \cite{Roxburgh2003a} proposed two other ratios, $r_{01}$ and $r_{13}$, as diagnostics of the interior of solar-like stars. $r_{01}$, defined as a $5$ point separation, captures fine details that the models cannot yet reproduce accurately. $r_{13}$ is defined in an analogous way to $r_{02}$, but taking the small separation between modes with $\ell=1$ and $3$. Unfortunately, due to partial cancellation \cite{Aerts2010}, we cannot yet detect $\ell=3$ modes, except for the Sun.

Numerical tests made by \cite{Roxburgh2003a} found that $r_{02}$ and similar ratios have an accuracy of $0.5 \%$, as explained by these authors the uncertainties in such ratios are solely due to the dependence of the inner phase shifts of the acoustic modes in the stellar structure. This high accuracy is possible because the influence of the problematic external layers of sun-like stars is suppressed in these ratios. Indeed, current stellar models have a poor description of the external layers of stars, with no inclusion of non-adiabatic convection and oscillations, and unrealistic stellar atmospheres. Hence, such ratios are a powerful tool to probe the core of stars when high quality data is available, like the data obtained by the space missions COROT and Kepler. Accordingly, we choose to use such ratios in this study to probe the core of our sun-like star targets, however, to be in the conservative side of the argument, we adopt an uncertainty of $1 \%$.

The discrepancies arising from the solar composition problem are the main source of uncertainty to our standard solar model. We captured the SSM uncertainty from the difference between a solar model computed with the AGSS09ph abundances and another computed with the GS98 composition, determining a $2 \%$ change in $r_{02}$ on average over the individual modes. We adopted this value as the reference uncertainty for all solar models. It is particularly important to have a conservative error estimate for the solar models because solar oscillation frequencies are determined very precisely. Otherwise, without the uncertainty from the solar model, we would be using a very precise diagnostic for a comparatively inaccurate model. Naturally, the model would be excluded due to the lack of detail in the analysis.

\section{Selecting and calibrating a star} \label{model}

\setlength{\tabcolsep}{5pt} 
\renewcommand{\arraystretch}{1} 

\begin{table*}
\begin{tabular}{rrrrrrr}
\toprule
Star & $Z$ & $T_{eff} (\text{K})$ & $Z/X$ & $\log(g/(\text{cm}/\text{s}^2))$ & $\Delta \nu (\mu \text{Hz})$ \\
\colrule
Sun & $0.0134$ & $5777$ & $0.0181$ & $4.438$ & $135$ \\
KIC 7871531 (Star A) & $0.0113 \pm 0.0054$ & $5400 \pm 100$ & $0.016 \pm 0.004$ & $4.4 \pm 0.2$ & $150 \pm 5$ \\
KIC 8379927 (Star B) & $0.0117 \pm 0.0054$ & $6000 \pm 200$ & $0.018 \pm 0.008$ & $4.4 \pm 0.2$ & $120 \pm 2$ \\
\botrule
\end{tabular}
\caption{Input observational constraints for the benchmark models. Solar values are shown for reference. \label{input}}
\end{table*}

\begin{table*}
\begin{tabular}{rrrrrrrrrr}
\toprule
Star & $M (\text{M}_{\astrosun})$ & $Z$ & $\tau (\text{Gyr})$ & $T_{eff} (\text{K})$ & $Z/X$ & $\log(g/(\text{cm}/\text{s}^2))$ & $R (\text{R}_{\astrosun})$ & $L (\text{L}_{\astrosun})$ & $\left\langle \Delta \nu \right\rangle (\mu \text{Hz})$ \\
\colrule
Sun & $1$ & $0.0134$ & $4.57$ & $5777$ & $0.0181$ & $4.438$ & $1$ & $1$ & $135$ \\
KIC 7871531 (Star A) & $0.85$ & $0.0140$ & $9.41$ & $5487$ & $0.0167$ & $4.47$ & $0.886$ & $0.639$ & $149.2$ \\
KIC 8379927 (Star B) & $1.12$ & $0.0180$ & $1.82$ & $6158$ & $0.0235$ & $4.39$ & $1.12$ & $1.62$ & $120.2$\\
\botrule
\end{tabular}
\caption{Resulting benchmark model parameters. Solar values are shown for reference. \label{output}}
\end{table*}

Undoubtedly our knowledge of the Sun renders it the best stellar laboratory for constraining DM. Nevertheless, many good laboratories are sometimes statistically more relevant than an excellent one.  \textit{Kepler} observed many stars showing a large number of detected oscillation modes \citep{Appourchaux2012}. From this set, interesting candidates for modelling have stellar fundamental properties strongly constrained by photometric and spectroscopic observations. Within that subset, the best candidates are those for which astrometric observations are also available, for example stars observed by both \textit{Kepler} and \textit{Hipparcos} \citep{SilvaAguirre2012}. Thus, the ideal candidate is an object with highly constrained fundamental properties and a large number of detected oscillation modes. Binary stars are also very interesting since some of their fundamental properties can be determined to high precision \citep{Torres2009}. Additionally, models of sub-solar mass stars can evidence a greater DM impact than models of stars more massive than the Sun. This is because in less massive stars, the DM luminosity corresponds to a greater fraction of the total luminosity, making these objects of particular interest.

In this work, we study the Sun to check the robustness of our models and to set a standard for our analysis. We also study a sub-solar mass star, KIC 7871531, of spectral type G5V and a modelled mass of $0.85 \, \text{M}_{\astrosun}$. KIC 7871531 (hereafter refereed to as star A) was observed by \textit{Kepler} and subsequently \cite{Appourchaux2012} identified $26$ oscillation modes for this star. Additionally, we also modelled KIC 8379927, a star of spectral type F9IV-V and a modelled mass of $1.12 \, \text{M}_{\astrosun}$. KIC 8379927 (hereafter refereed to as star B) was observed by both \textit{Kepler} and \textit{Hipparcos} and $37$ modes have been detected by \citep{Appourchaux2012}.

It is possible to calibrate the solar model to precisely match the well known solar fundamental parameters. Moreover, the solar age is determined with considerable precision, for example from meteoritic analysis \citep{Connelly2012}. On the contrary, for other stars even if the mass and metallicity were known with an acceptable precision, the age would not. Consequently, there is a degeneracy in the fundamental parameters which must also be considered, even before the effects of DM come into play.

To find a calibrated stellar model for stars other than the Sun we proceeded by first building a set models of those stars within a grid of masses, metallicities and ages, $(M,Z/X,\tau)$ without including the effects of DM. We then compare each model within that set with input observational constraints and pick only a subset which satisfies those restrictions. For each star, we considered input observational constraints on the effective temperature $T_{eff}$, total iron content $[Fe/H]$, surface gravity $g$ and on the average of the large separation $\langle \Delta \nu \rangle$. For all stars, $\langle \Delta \nu \rangle$ was computed over all the possible $n$, for $l=0$ only. A different number of large separations was used to compute the average for each star, $22$ large separations for the Sun (with $n=7$ through $28$), $6$ for star A ($n=19$ through $24$), and $12$ for star B ($n=16$ through $27$). The $\langle \Delta \nu \rangle$ constraint was compared against the primary frequency splitting obtained from the scaling relation $\Delta \nu_{0} = (M/M_{\astrosun})^{1/2} (R/R_{\astrosun})^{-3/2} \, 134.9 \, \mu \text{Hz}$ \cite{Kjeldsen1995}, thus further constraining the models in the initial grid, even if they satisfy the other constraints, including the surface gravity $g$, which also relates the mass and radius of the star. These constraints given in table \ref{input} are based on \cite{Bruntt2012a} and \cite{Molenda-Zakowicz2013} for star A and on \cite{Karoff2013b} and \cite{Molenda-Zakowicz2013} for star B. We then compute the asteroseismic diagnostics for the subset satisfying the input observational constraints of models and compare those diagnostics to the results inferred from observation. The closest model, which best mimics the observed star, in particular in the core, is the benchmark model. It is against this model that we compare any model of this star including DM. Table \ref{output} shows the resulting parameters for the benchmarks, which are similar to those found throughout the literature, see for example \cite{Metcalfe2014a} for star A and \cite{Mathur2012a} and \cite{Karoff2013b} for star B. The DM stellar models of this star are build by taking the benchmark model and including the effects of DM.

\section{Results} \label{results}

In total, more than $1000$ CPU hours (@ $2.93 \text{GHz}$) were used to compute the stellar models corresponding to the results presented here.

\begin{figure}

\subfloat{\includegraphics[clip=true, trim= 0.5cm 0cm 3cm 1.7cm, width=1\columnwidth]{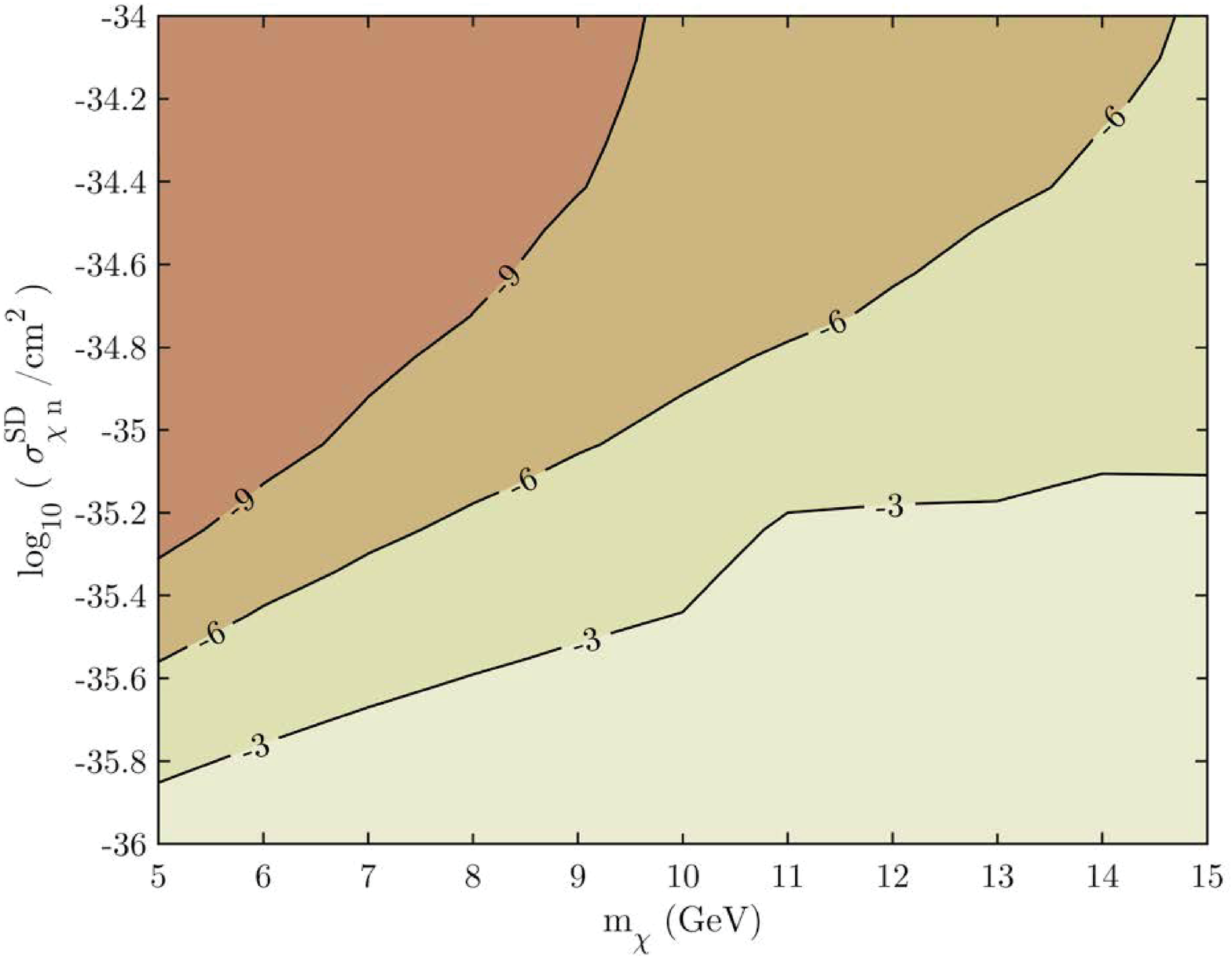} \label{sun_adm_tc_pct}}

\subfloat{\includegraphics[clip=true, trim= 0.5cm 0cm 3cm 1.7cm, width=1\columnwidth]{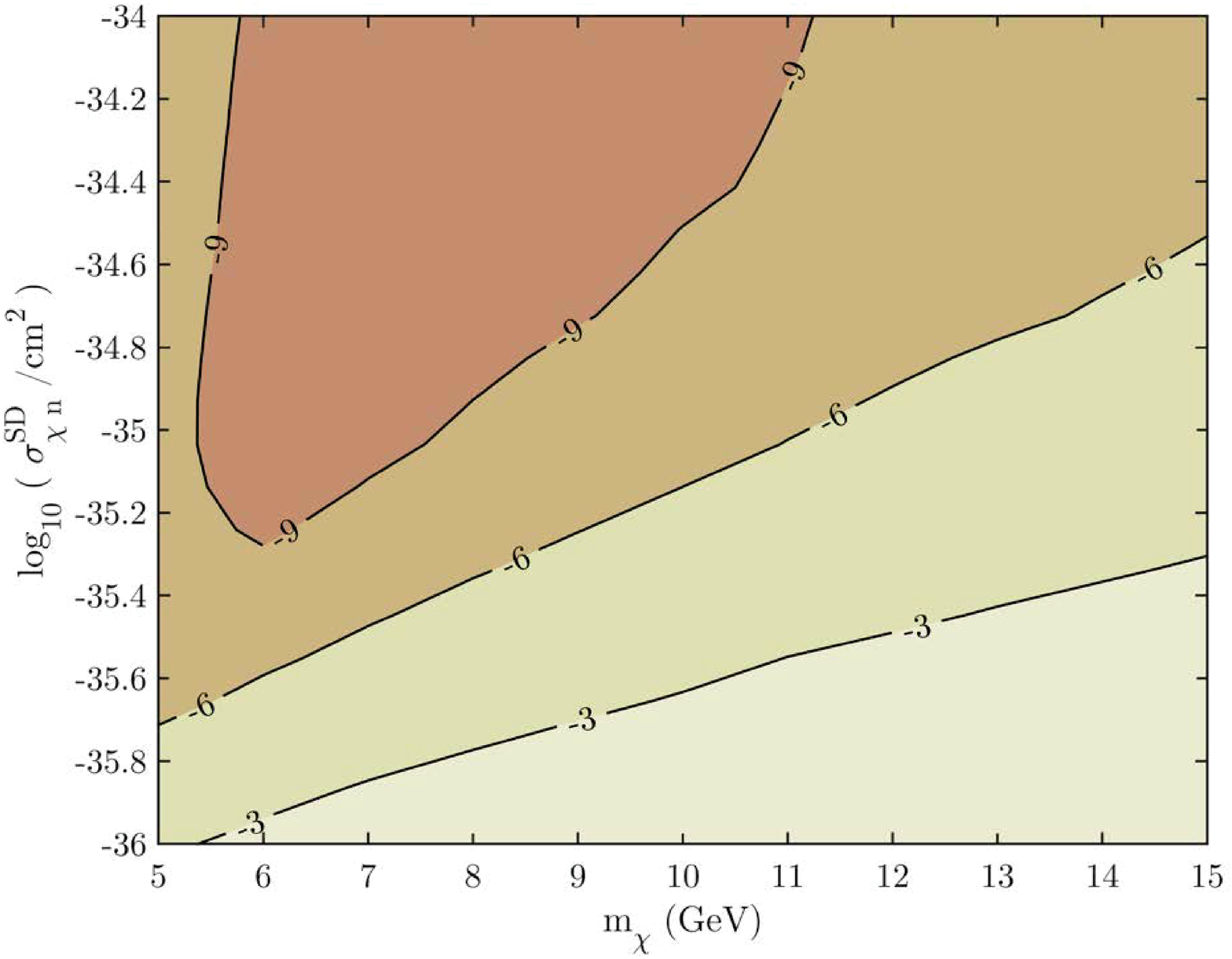} \label{kic7871531_adm_tc}}

\subfloat{\includegraphics[clip=true, trim= 0.5cm 0cm 3cm 1.7cm, width=1\columnwidth]{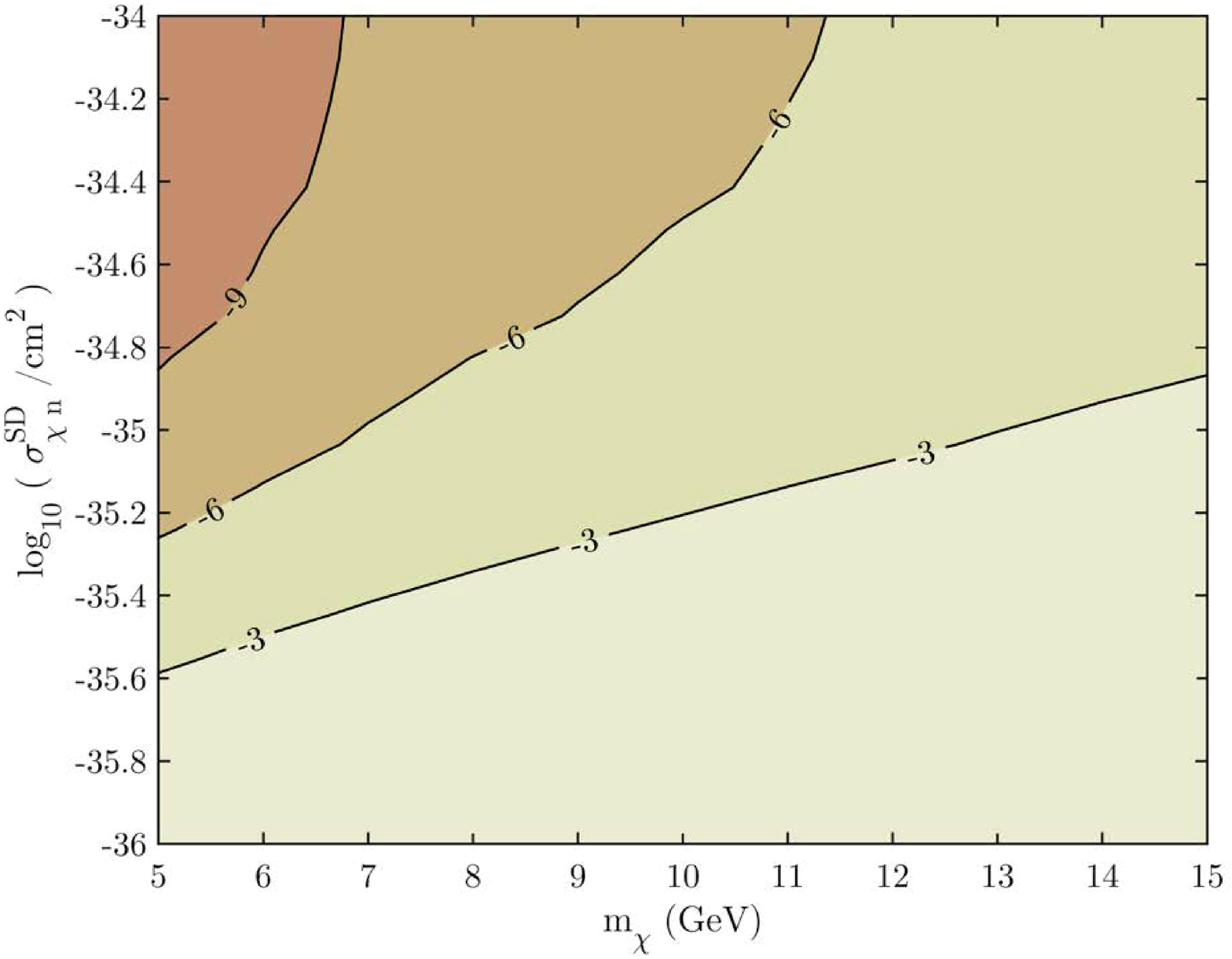} \label{kic8379927_adm_tc}}

\caption{Relative differences in the central temperature of models including ADM ($\delta T_{c}/T_{c}^{bench}$ in \%) for the Sun (top, $T_{c}^{bench} = 15.46 \, \text{MK}$), the less massive star A, KIC 7871531 (middle, $T_{c}^{bench} = 15.2 \, \text{MK}$) and the more massive star B, KIC 8379927 (bottom, $T_{c}^{bench} = 16.6 \, \text{MK}$). \label{adm_tc}}
\end{figure}

\subsection{Sun}

\begin{figure*}
\subfloat{\includegraphics[clip=true, trim= 0.5cm 0cm 3cm 1.7cm, width=1\columnwidth]{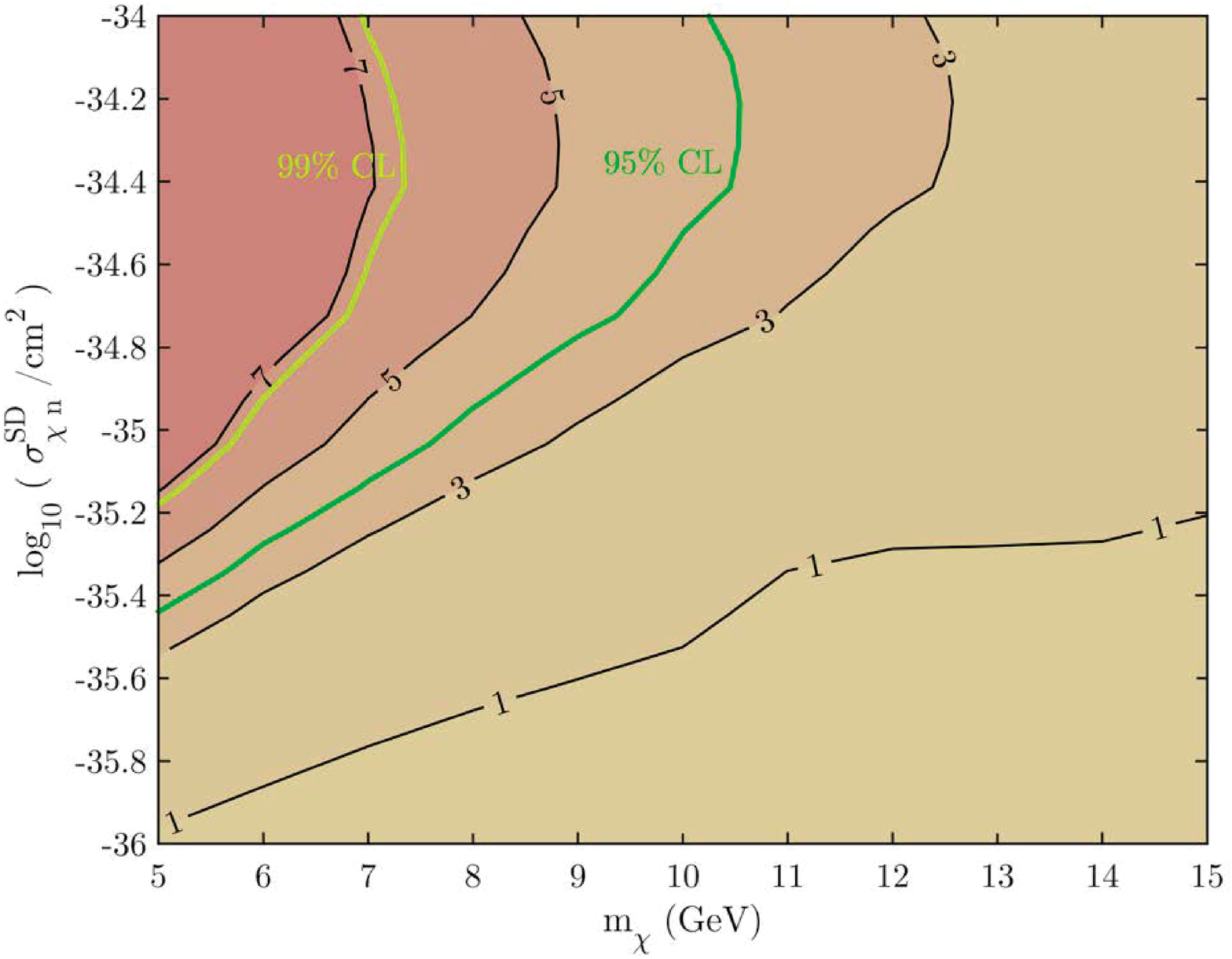}} \label{sun_wimp_tc}
~
\subfloat{\includegraphics[clip=true, trim= 0.5cm 0cm 3cm 1.7cm, width=1\columnwidth]{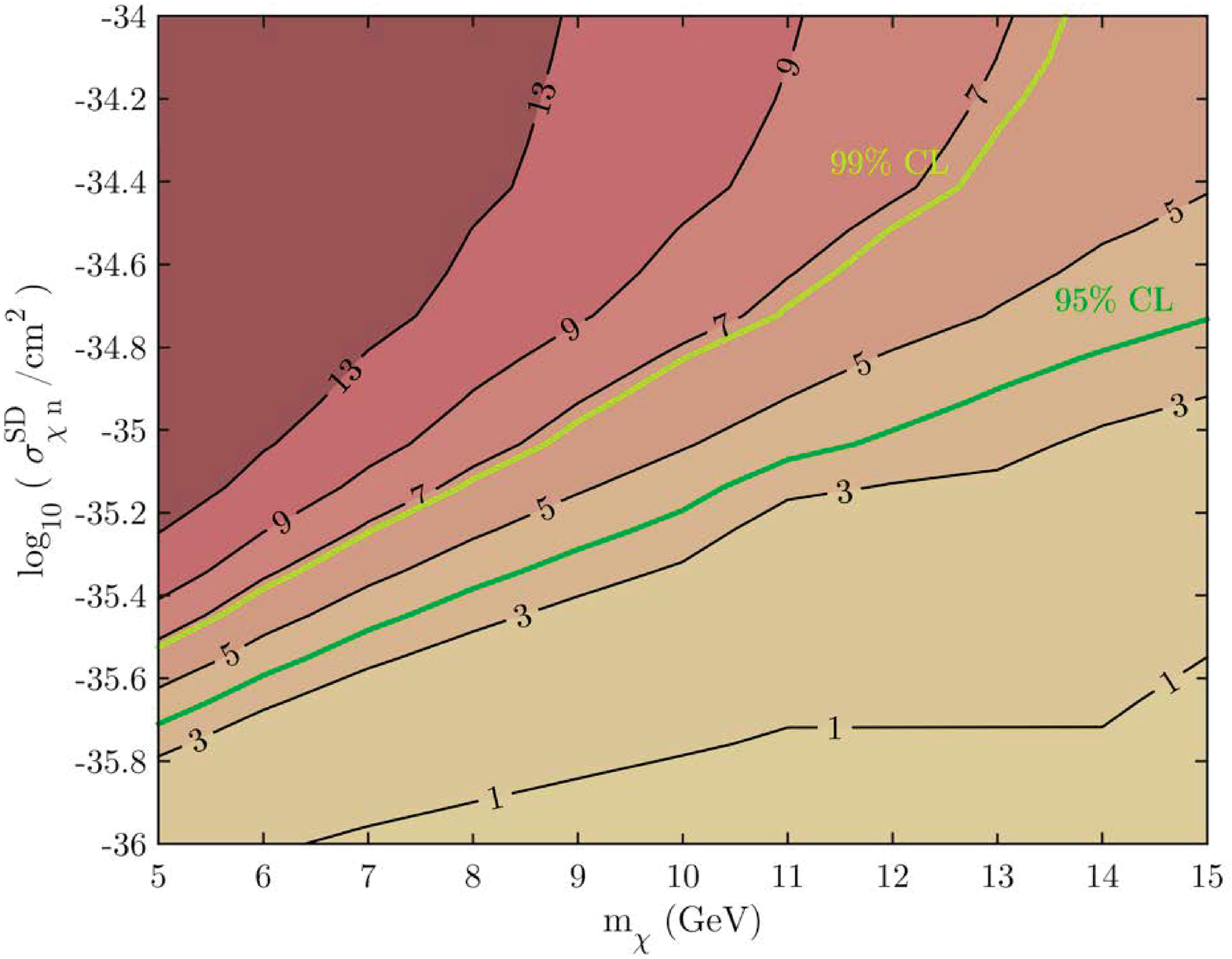}} \label{sun_adm_tc}
\caption{Squared errors ($\chi_{T_{c}}^{2}$) for the central temperatures of solar models including WIMPs (left) and ADM (right).  \\ The reference SSM central temperature $T_{c} = 15.750 \pm 0.5 \, \text{MK}$ is from a seismic solar model which stabilized the neutrino flux and gravity mode frequency predictions given by solar models \cite{Turck-Chieze2011}. Also shown are the $95\%$ and $99\%$ CLs. \label{sun_dm_tc}}
\end{figure*}

ADM has a considerable impact on the Sun. Figure \ref{sun_adm_tc_pct} shows the drop in central temperature, relative to the benchmark, for models including ADM. In the low mass, high cross section region of the parameter space, the drop due to ADM reaches $\sim 12 \%$ of the benchmark central temperature. This is in contrast with WIMPs, for which the drop does not go beyond $\sim 7 \%$, even at a very low annihilation rate $\left\langle \sigma_{A} v \right\rangle \sim 10^{-33} \, \text{cm}^{3}/\text{s}$. ADM can have a considerable greater impact than WIMPs, without the need to push the annihilation rate to extremely low values.

To further illustrate this point, we compared the differences between the two scenarios taking the SSM central temperature to be $T_{c}^{SSM} = 15.750 \pm 0.5 MK$ \citep{Turck-Chieze2011}. This central temperature is merely a reference and it corresponds to the central temperature of a seismic solar model which stabilized the neutrino flux and gravity mode frequency predictions given by solar models. An analysis solely based on this value already disfavours some regions of the parameter space as shown in figure \ref{sun_dm_tc}, where
\begin{equation}
\chi_{T_{c}}^{2} = \left( \frac{T_{c}^{SSM}-T_{c}^{mod}}{\sigma_{T_{c}}^{SSM}} \right)^{2} .
\end{equation}
For a particle of mass $m_{\chi} = 5 \, \text{GeV}$ for example, WIMPs with $\sigma^{SD}_{\chi n} \gtrsim 6.3 \times 10^{-36} \, \text{cm}^{2}$ are disfavoured, as well as ADM particles with $\sigma^{SD}_{\chi n} \gtrsim 3.1 \times 10^{-36} \, \text{cm}^{2}$, up to a $99\%$ confidence level (CL). This emphasizes the potential of a solar central temperature diagnostic for ADM, provided that the precision of solar neutrino measurements increases, together with the accuracy of SSM.

Although a solar central temperature analysis seems encouraging for the prospect of DM searches, asteroseismology arises as even more promising. To frame the asteroseismic analysis for stars A and B, we first compared the $r_{02}$ diagnostics obtained from helioseismic observational data and those determined from the computed solar models including ADM. We found that the low-mass, high-cross section region of the parameter space shows a stark disagreement not only with the observational data, but also with the benchmark. In fact, in that region, the discrepancy is much more significant than some degree of incompatibility found between the benchmark and observational data, which was to be expected due to model inaccuracies. This confirms that the low-mass, high-cross section region of the parameter space is indeed significantly disfavoured. Again, using the example of a $5 \, \text{GeV}$ particle, ADM with $\sigma^{SD}_{\chi n} \gtrsim 6.3 \times 10^{-36} \, \text{cm}^{2}$ is incompatible with the $r_{02}$ data obtained from helioseismology at least at a $99 \%$ CL.

\subsection{KIC 7871531 (star A, $0.85 \text{M}_{\astrosun}$)}

\begin{figure}
\includegraphics[clip=true, trim= 0.5cm 0cm 3cm 1.7cm, width=1\columnwidth]{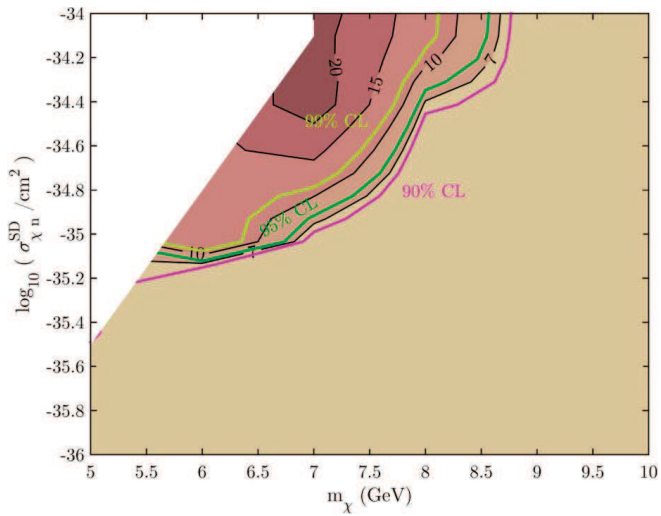}
\caption{Sum of squared errors for the $r_{02}$ diagnostic of KIC 7871531 (star A) models including ADM particles with $\sigma_{\chi \chi} = 10^{-24} \, \text{cm}^{2}$. Also shown are the $90\%$, $95\%$ and $99\%$ CLs corresponding to these $\chi^{2}$s with the number of $d.o.f. = 3$. The empty region in the low mass, high cross section region of the parameter space corresponds to stellar models that did not converge. \label{kic7871531_adm_r02}}
\end{figure}

The benchmark model obtained for star A has a mass of $0.85 \text{M}_{\astrosun}$ and an age of $9.41\text{Gyr}$. Thus, we expected to find a greater DM impact for this star than for the Sun, not only because this is a less massive star, for which the energy transported by DM represents a larger fraction of the total luminosity, but also because DM would have accumulated for longer inside this star. The results shown in figure \ref{kic7871531_adm_tc}, compared with those in figure \ref{sun_adm_tc_pct}, confirm this expectation for ADM. In fact, both for WIMPs and for ADM, the drop in central temperature, relative to the benchmark, is greater for star A than for the Sun, almost everywhere in the parameter space. Take an ADM particle with $m_{\chi} = 10 \, \text{GeV}$ and $\sigma^{SD}_{\chi n} = 10^{-35} \,  \text{cm}^{2}$, while for the Sun the drop is just short of $6 \%$ of the benchmark central temperature, for star A, the enhanced energy transport implies a reduction of almost $8 \%$.

For stars other than the Sun we do not have central temperature estimates based on neutrino observations, hence we cannot conduct a central temperature analysis as we did for the Sun. Nonetheless, asteroseismology offers a probe of the stellar core. A $\chi_{r_{02}}^{2}$ diagnostic evidences the disagreement between asteroseismic data and models including ADM as shown in figure \ref{kic7871531_adm_r02}. The stellar models on the top left of figure \ref{kic7871531_adm_r02} did not converge because the atmosphere could not be reconstituted. In a similar trend as the Sun, the low-mass, high-cross section region of the parameter space is clearly incompatible with the asteroseismic data. For that region, there is also a significant departure from the benchmark, which is not present throughout the rest of the parameter space. An ADM particle with $m_{\chi} = 5 \, \text{GeV}$ and $\sigma^{SD}_{\chi n} = 4 \times 10^{-36} \,  \text{cm}^{2}$ is strongly disfavoured by our analysis, although this limit on the cross section increases to considerable less stringent values for heavier particles, with masses $\geq 6 \, \text{GeV}$. A comparison between the $\chi_{r_{02}}^{2}$ results of star A and of solar models including ADM also confirms the expectation that this less massive, older star evidences a greater DM impact. While the Sun disfavours ADM particles with smaller cross sections below $6 \, \text{GeV}$, star A is more competitive above that mass. In fact, whereas the Sun excludes ADM with $\sigma^{SD}_{\chi n} \lesssim 10^{-34} \,  \text{cm}^{2}$ for masses just shy of $8 \, \text{GeV}$ at $99 \%$ CL, this star can go up to almost $9 \, \text{GeV}$. This is interesting considering that we are matching the Sun, outfitted with $17$ individual $r_{02}$ ratios, against a star for which only $3$ of these ratios are available.

\subsection{KIC 8379927 (star B, $1.12 \, M_{\astrosun}$)}

\begin{figure*}

\subfloat{\includegraphics[clip=true, trim= 0.5cm 0cm 3cm 1.7cm, width=1\columnwidth]{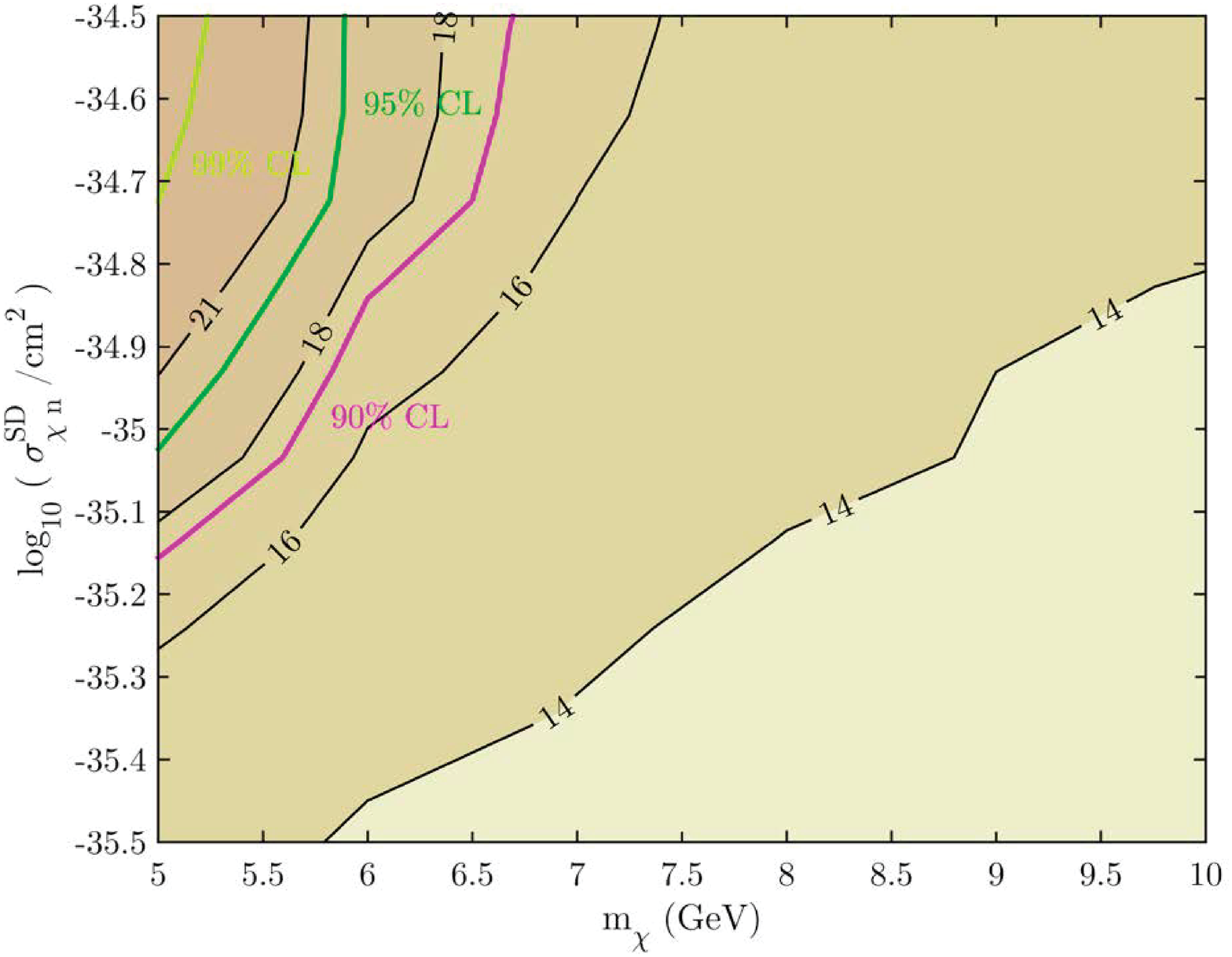} \label{kic8379927_wimp_r02}}
~
\subfloat{\includegraphics[clip=true, trim= 0.5cm 0cm 3cm 1.7cm, width=1\columnwidth]{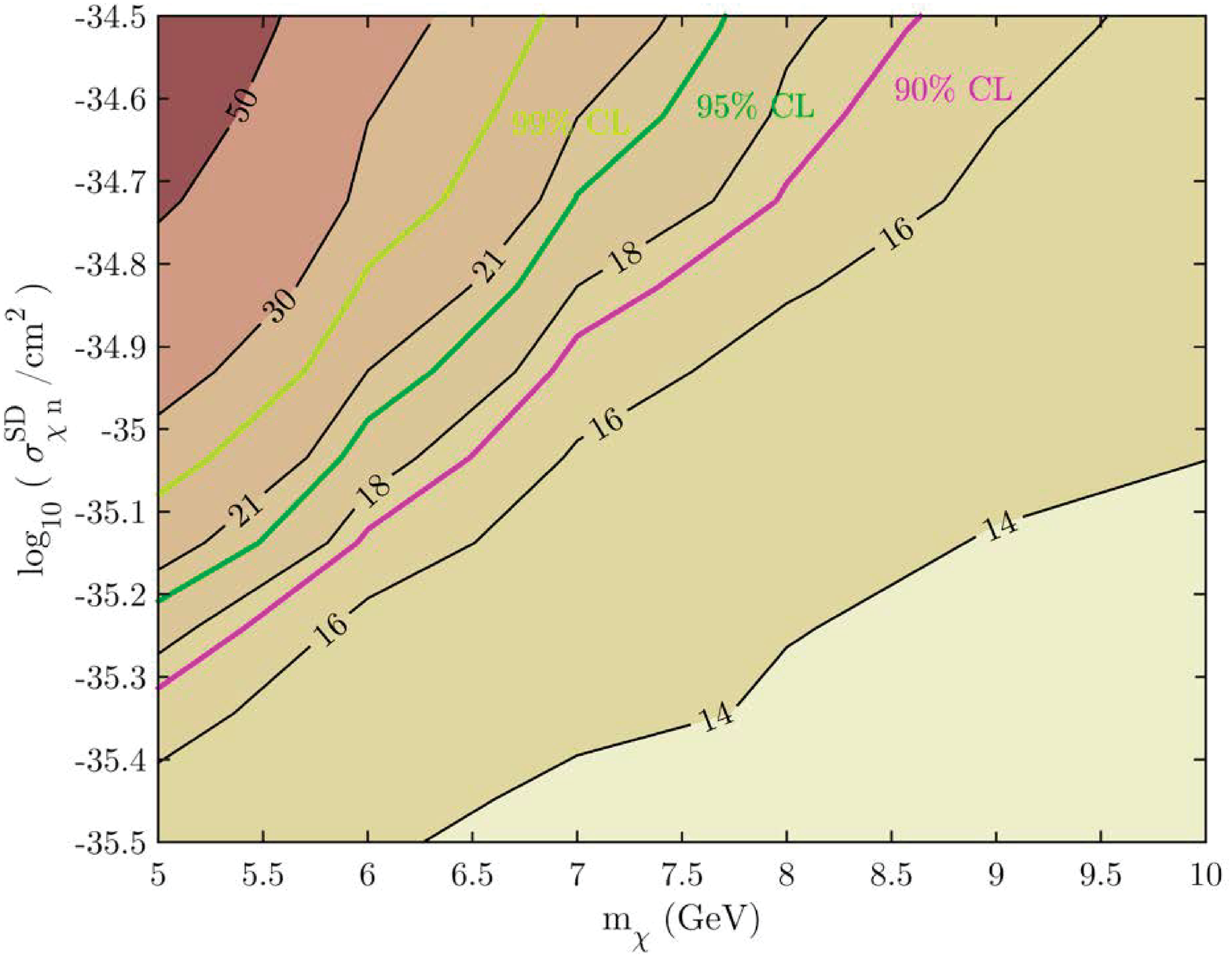} \label{kic8379927_adm_r02}}

\caption{Sum of squared errors for the $r_{02}$ diagnostic of star B (KIC 8379927) models including WIMPs (left) and ADM (right).
	Also shown are the $90\%$, $95\%$ and $99\%$ CLs corresponding to these $\chi^{2}$s with the number of $dof = 11$. \label{kic8379927_dm_r02}}
\end{figure*}

Star B has a benchmark model slightly more massive than the Sun, with $1.12 \, M_{\astrosun}$ and also considerably younger, with $1.82 \, \text{Gyr}$. As previously discussed, in general, we would expect DM to produce a smaller departure from standard stellar modelling for a star like this. Again, this hypothesis is supported by the comparison between the drop in central temperature for models of the more massive star B and of the Sun, figures \ref{kic8379927_adm_tc} and \ref{sun_adm_tc_pct}, respectively.

For star B, $11$ individual $r_{02}$ ratios are available. This is considerably more than the $3$ ratios accessible for star A, but still less than the $17$ available for the Sun. Also, on average, the precision of an $r_{02}$ ratio for star B is $\sim 15 \%$ better than for A. We computed the $\chi_{r_{02}}^{2}$ diagnostic for models of star B with WIMPs and ADM, the results are shown in figure \ref{kic8379927_dm_r02}. 

ADM impact is most significant for this star for larger DM masses since a higher core temperature leads to a mean free path of the DM particle $\sim 10 \%$ greater than that of the less massive star A for that mass range. As a consequence, energy transport due to a single DM particle is more efficient in this case. We note however that the total energy transported by DM depends both on the number of accumulated particles and the efficiency of the transport. For a $5 \, \text{GeV}$ particle, WIMPs with $\sigma^{SD}_{\chi n} = 6.3 \times 10^{-36} \,  \text{cm}^{2}$ and ADM with $\sigma^{SD}_{\chi n} = 4.5 \times 10^{-36} \,  \text{cm}^{2}$ are incompatible with the observational data up to a $99 \%$ CL.

\section{Discussion \label{discussion_conclusions}}

\begin{figure*}

\subfloat{\includegraphics[clip=true, trim= 0.5cm 0cm 3cm 1.5cm, width=1\columnwidth]{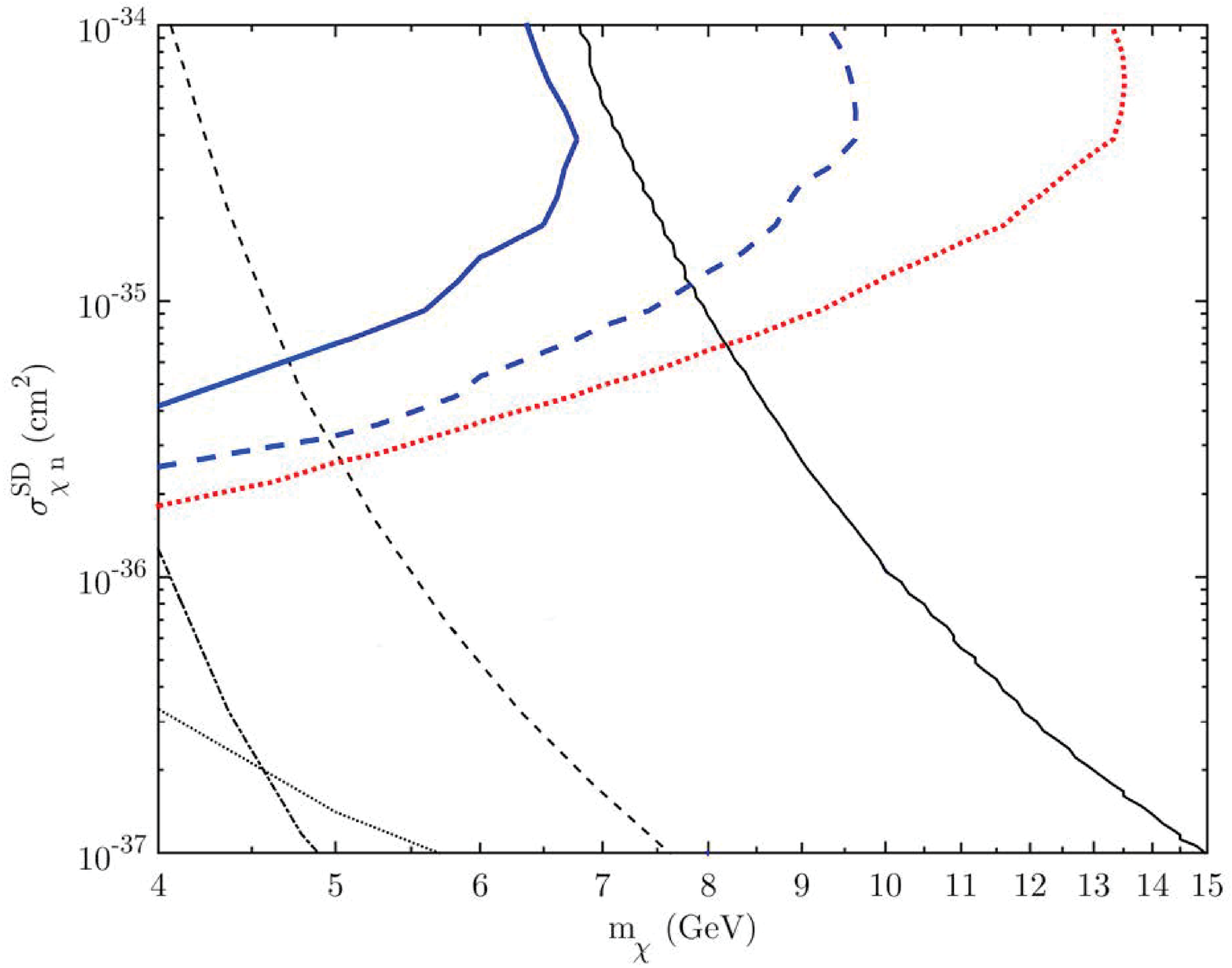} \label{limits_WIMP}}
~
\subfloat{\includegraphics[clip=true, trim= 0.5cm 0cm 3cm 1.5cm, width=1\columnwidth]{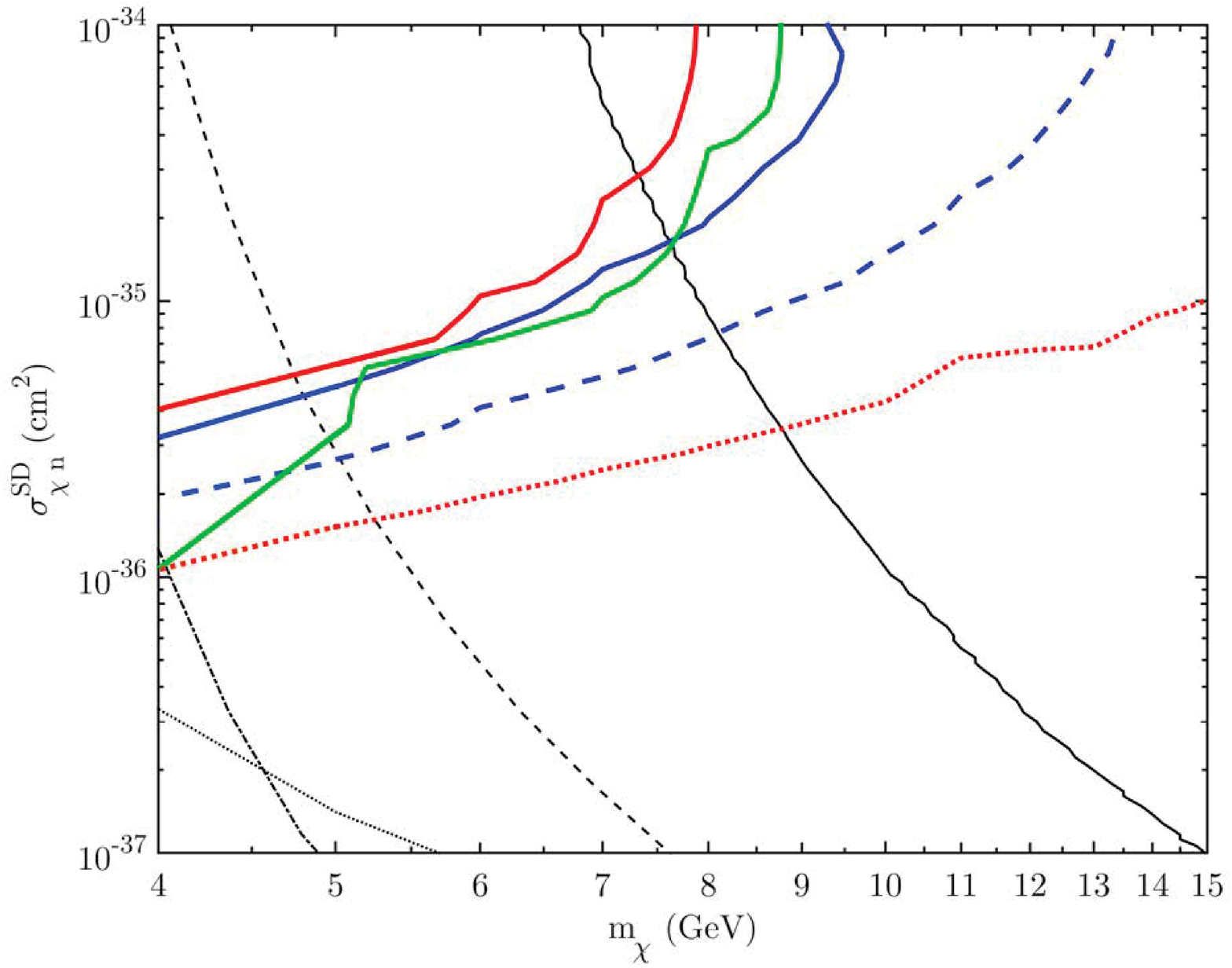} \label{limits_ADM}}

\caption{$90\%$ CL limits ascertained from this work in the WIMP (left) and ADM (right) scenarios for: Sun $\chi_{T_{c}}^{2}$ (dotted {\color{red} red}), Sun $\chi_{r_{02}}^{2}$ (solid {\color{red} red}), star B (KIC 8379927) $\chi_{r_{02}}^{2}$ (solid {\color{blue} blue}), star A (KIC 7871531) $\chi_{r_{02}}^{2}$ (solid {\color{green} green}). The dashed {\color{blue} blue} line is the projected $90\%$ CL limit corresponding to a $10\%$ increase in precision for the mode frequencies. The impact due to WIMPs could only be significantly diagnosed for star B (KIC 8379927), hence the absence of constraints for the other stars. For comparison, $90\%$ CL limits from some direct detection experiments are also shown (references given in text): XENON100 (solid black line), COUPP flat efficiency model (dashed black line), PICO-2L (dash-dotted black line), PICASSO (dotted black line). \label{text}}
\end{figure*}

In an effort to understand the potential of asteroseismology as a complementary way to search for DM, we analysed the effects of WIMPs and ADM in three stars using the core-sensitive asteroseismic ratio $r_{02}$. As such, we attempted to constrain the properties of low-mass ADM with an effective spin-dependent coupling by comparing observational data with results of stellar models including DM energy transport.

The asteroseismic analysis disfavours the low mass, high cross section region of the ADM parameter space explored here. This incompatibility was found for all three stars. For example, at $99 \, \%$ CL, for $m_{\chi} = 5 \, \text{GeV}$, the $r_{02}$ data is incompatible with ADM models with $\sigma_{\chi n}^{SD} \gtrsim 6 \times 10^{-36} \text{cm}^{2}$ for the Sun, $\sigma_{\chi n}^{SD} \gtrsim 5 \times 10^{-36} \text{cm}^{2}$ for star A and $\sigma_{\chi n}^{SD} \gtrsim 4 \times 10^{-36} \text{cm}^{2}$ for star B.

In this work, we considered only an effective SD coupling for the DM-nucleon interaction. Our results for the Sun show less dramatic differences between data and model than what \cite{Vincent2015a} found. This can be explained by the additional reference uncertainty, which we included to reflect the unsolved solar abundance problem. One way to look at the problem is to take the most recent abundances of AGSS09ph at face value. The other, which we adopted here, is not so much to question the accuracy of those abundances, but instead to capture some of the uncertainties in the SSM by considering the differences between solar models computed with the abundances by AGSS09ph and those by GS98.

Figures \ref{limits_WIMP} and \ref{limits_ADM} show the limits and exclusion regions ascertained from this work for the WIMP and ADM scenarios, respectively. For comparison, limits are also shown for some direct detection experiments: the scintillation and ionization detector XENON100 \citep{Aprile2013} and the bubble chambers COUPP \citep{Behnke2012}, PICASSO \citep{Archambault2012} and PICO-2L \citep{Amole2015}.

In the case of WIMPs, where accumulation is considerably weaker, only the more massive star B is significantly affected. Both the Sun and star A are relatively unaffected by WIMPs within the parameter range explored here, even though they are older stars. There is a trade-off between DM accumulation and stellar age in terms of DM impact.

Notice from our solar central temperature analysis that solar models with low-mass, high-cross section ADM do not compare well with current SSM. This comparison should be taken cautiously since the central temperature of SSM is very sensitive to model input physics. The SSM has been refined over the last two decades to account for the solutions put forward to solve several problems in solar physics \citep{Turck-Chieze2011}, but a definite answer is yet to be found. More accurate abundances and opacities, as well as more accurate and precise measurements of the $^{8}B$ neutrino flux could alter the prediction for the central temperature of the SSM.

For $m_{\chi} \gtrsim 5 \, \text{GeV}$, the best constraints from the $r_{02}$ analysis come from star B, a star slightly more massive and considerably hotter, more metallic and younger than the Sun. For ADM, the $90 \%$ CL limit set by this star is competitive with the $90 \%$ CL bounds set by XENON100 for masses below $\sim 7 \, \text{GeV}$. For $m_{\chi} \lesssim 5 \, \text{GeV}$, star A gives the most stringent bounds, competitive with the $90 \%$ CL limits set by COUPP. The ADM impact is so significant in this case because this is an older star, with slightly more than twice the age of the Sun. Because we are dealing with stellar evolution, the effects of DM energy transport are cumulative, which means that significantly older stars can be competitive in constraining DM. In fact, while this star would not show a significant DM impact at a younger age, it does so at an older one.

An interesting aspect of using asteroseismology to constrain DM resides in the fact that objects with different fundamental properties can be affected at distinct levels. We have already discussed the case of less massive stars which exhibit a greater DM impact. Moreover, we have also mentioned that older stars are more affected than similar younger objects, which is only natural since DM accumulated for longer and to a greater number in those older stars. The number of ADM particles grows linearly with time after the geometrical limit is reached early on in the life of a star and the luminosity transported by DM is proportional to that number. Hence, the luminosity grows linearly with the age of the star. For an object like star A, with an age roughly twice that of the Sun, we can expect the number of accumulated particles to be greater by a factor of $2$ in comparison with the Sun. This can translate into a significant impact on the stellar structure which explains, at least in part, why we were able to set constraints using star A, for which only $3$ individual $r_{02}$ ratios are available. In the future, stellar age should also be a relevant selection criteria to look for when attempting to constrain DM properties with asteroseismology.

We asked whether these ADM particles significantly alter the star's internal structure during the course of the star's evolution. We conclude that they do. Moreover, we found that ADM models can be excluded using astroseismic data with a high statistical significance, for the regions of the parameter space presented before. However, we would like to draw attention to a couple of caveats in our analysis. First, we recall that the solar composition problem indicates some missing physics in the SSM which most definitely affect the results. We accounted for this uncertainty when comparing between solar model and observational data by considering a reference uncertainty corresponding to the difference between models computed with AGSS09ph and GS98. This is not an issue for the other two stars, for which the frequencies are determined with a worse precision, that covers the reference uncertainty. Second, we note that the calibration processes are prone to degeneracies in the stellar parameters. We would not be able to significantly disfavour a particular DM model if a small change in the benchmark parameters of the stellar model gave a very different result by bringing the $r_{02}$ from the model, closer to that of the observational data. We circumvented this issue by making sure that for a small change in the benchmark parameters of the stellar model there is only a small change in the diagnostic, for a DM model with maximal impact.

It is possible that a more sophisticated interaction such as a $q^{2}$ momentum-dependent cross sections would be favoured by a similar analysis at lower cross section values. It would be very interesting to study star B considering instead a momentum-dependent coupling as \cite{Vincent2015a} did for the Sun. For this it will be necessary to review the mass evaporation threshold and to understand whether different stars allow for different limits.

The Transiting Exoplanet Survey Satellite (TESS) and the PLAnetary Transits and Oscillations of stars (PLATO) mission are expected to allow for the determination of mode frequencies with even better precisions than \textit{Kepler}. Figure \ref{limits_ADM} displays the projected $90\%$ CL ADM limit for the asteroseismic analysis of star B with a $10 \%$ increase in the frequency precision of all modes. Furthermore, improvements of more than $10\%$ seem feasible, for example, a frequency precision of about $0.1 \mu \text{Hz}$ is expected to be achieved with PLATO \citep{Rauer2014}, corresponding to an average improvement of at least $\sim 30 \%$ across all the detected modes of star B. 

Moreover, \textit{Gaia}'s first Intermediate Data Release is expected in a few months and  it will already provide parallaxes for common \textit{Kepler} targets. In the future, about one billion stars will be mapped and naturally Gaia will share quite a few targets with TESS and PLATO. This raises the possibility of having a large number of stars with frequencies determined to very high precision for which high-quality astrometric data is also available, thus setting tighter input observational constraints for mode ling attempts.

As our understanding of stellar physics improves, asteroseismic diagnostics are starting to offer a complementary approach to corroborate direct detection limits. Asteroseimic studies of DM are essentially competitive for low DM masses of a few GeV, just above the evaporation mass, where DM produces significant impact in stars. This is a very interesting result considering that the present direct DM detection experiments are not able to accurately probe the parameter space of low mass DM particles.

\acknowledgments
JC acknowledges the support of the Alexander von Humboldt-Stiftung Foundation. We thank the referee for his/her comments, which significantly improved our work.


\end{document}